\documentclass[11pt]{article}
\usepackage{amsmath}
\usepackage{latexsym}
\usepackage{amssymb}
\begin{document}

\title{\bf{\Large{Quantum entanglement in multiparticle systems of two-level atoms}}}
\author{\bf{\normalsize Ram Narayan Deb}
\thanks{debram@rediffmail.com}\\ 
Department of Physics, Krishnagar Government College,\\
 Krishnagar, Nadia, Pin 741101, West Bengal, India}
\date{}

\maketitle

\begin{abstract}
We propose the necessary and sufficient condition for the presence of
quantum entanglement in arbitrary symmetric pure states of two-level atomic systems. We introduce a parameter to quantify quantum entanglement in such systems. We express the inherent quantum fluctuations of a composite system of two-level atoms as a sum of the quantum fluctuations of the individual constituent atoms and their
 correlation terms. This helps to separate out and 
study solely the quantum correlations among the atoms
 and obtain the criterion for the presence of entanglement in such multiatomic systems. 
\end{abstract}
\maketitle
\section{Introduction}
Over the past few years there has been a growing interest in quantum
mechanically correlated multiatomic systems 
\cite{Hald}-\cite{Julsgaard2}.
Quantum entanglement, which is the basic ingredient of quantum information theory, is yet to be understood completely in the context of such systems. The proposal of the criterion made by Peres and Horodecki in Refs.
\cite{Peres} and \cite{Horodecki}, regarding the presence of quantum entanglement in a quantum system, forms
an important step toward the understanding of quantum entanglement
in the context of bipartite states.  It has also been found that spin squeezing has a close relationship with quantum entanglement and a lot
of work has been done in this direction 
\cite{Sorensen}-\cite{Toth}. A system which is in a spin squeezed state is also quantum mechanically entangled. But, quantum  entanglement does not ensure spin squeezing. A system which is in a  quantum mechanically
entangled state may not show spin squeezing always. Therefore, spin
squeezing cannot be used always to detect and quantify quantum entanglement. 

In this paper, we introduce
the necessary and sufficient condition for the presence of quantum
entanglement in multiatomic systems and also introduce a parameter to quantify quantum entanglement.

An atom has many energy levels, but when it is interacting with an external monochromatic electromagnetic field, we concentrate mainly on two of its energy levels among which the transition of the atom takes place. Hence, 
 the atom is called a two-level atom.

We consider a system of $N$ such two-level atoms. 
If among the two energy levels of the $n$-th atom in the assembly,
 the upper and lower energy levels 
are denoted as
$|u_n\rangle$ and $|l_n\rangle$, respectively, then we can construct a vector 
operator $\hat{\mathbf J}_n$ whose
components are  
\begin{eqnarray}
\hat{J}_{n_x} &=& (1/2)\big(|u_n\rangle\langle l_n| + |l_n\rangle
\langle u_n|\big),\label{1.1a1}\\
\hat{J}_{n_y} &=& (-i/2)\big(|u_n\rangle\langle l_n| - |l_n\rangle
\langle u_n|\big),\label{1.1a2}\\
\hat{J}_{n_z} &=& 
(1/2)\big(|u_n\rangle\langle u_n| - |l_n\rangle\langle l_n|
\big),
\label{1.1a3}
\end{eqnarray}
such that 
\begin{equation}
[\hat{J}_{n_x} , \hat{J}_{n_y}] = i\hat{J}_{n_z}
\label{1.1a4}
\end{equation}
and two more relations with cyclic 
changes in $x$, $y$, and $z$. 
Since the operators $\hat{J}_{n_x}$,
$\hat{J}_{n_y}$, and $\hat{J}_{n_z}$ obey the same commutation 
relations as the spin operators, these are called pseudo-spin operators.

For the entire system of $N$ two-level atoms, we
construct collective pseudo-spin operators 
\begin{eqnarray}
\hat{J}_x = \sum_{i=1}^{N}\hat{J}_{i_{x}},~~~~ 
\hat{J}_y = \sum_{i=1}^{N}\hat{J}_{i_{y}},~~~~ 
\hat{J}_z = \sum_{i=1}^{N}\hat{J}_{i_{z}},
\label{1.1a}
\end{eqnarray}
where it is implicitly assumed that each term in the 
above summations is in direct product with the identity
operators of all other atoms.

The individual atomic operators satisfy
\begin{eqnarray}
\big[\hat{J}_{1_x}, \hat{J}_{2_y}\big] = 0,~
\big[\hat{J}_{1_x}, \hat{J}_{1_y}\big] = i\hat{J}_{1_z}, ~
\big[\hat{J}_{2_x}, \hat{J}_{2_y}\big] = i\hat{J}_{2_z},...
\label{1.2a1}
\end{eqnarray}
 As a direct consequence of these commutation
relations we have
\begin{equation}
[\hat{J}_x , \hat{J}_y] = i\hat{J}_z
\label{1.2a2}
\end{equation} 
and two more
relations with cyclic changes in $x$, $y$, and $z$.

The simultaneous eigenvectors of $\hat{J}^2 = \hat{J}_x^2
+ \hat{J}_y^2 + \hat{J}_z^2$ and $\hat{J}_z$ are denoted
as $|j,m\rangle$, where
\begin{equation} 
\hat{J}^2|j,m\rangle = j(j+1)|j,m\rangle
\label{1.1}
\end{equation}
 and 
\begin{equation}
\hat{J}_z|j,m\rangle = m|j,m\rangle.
\label{1.2}
\end{equation} 
Here $j = N/2$ and $m = -j, -j+1, ....(j-1), j$.
The collective quantum state vector for a system of $N$
two-level atoms can be expressed as a linear superposition of $|j,m\rangle$ as  
\begin{equation}
|\psi_j\rangle = \sum_{m=-j}^{j}c_{m}|j,m\rangle.
\label{1.3}
\end{equation}
To find out whether a quantum state $|\psi_j\rangle$ of the system is an atomic coherent state \cite{Radcliffe} or an atomic squeezed state 
\cite{Kitagawa}, \cite{Wineland} we calculate the mean pseudo-spin
vector
\begin{equation}
\langle\hat{\mathbf{J}}\rangle = \langle\hat{J}_x\rangle
\hat{i} + \langle\hat{J}_y\rangle\hat{j} + \langle\hat{J}_z\rangle\hat{k}
\label{1.6}
\end{equation}
for the quantum state $|\psi_j\rangle$. The vector 
$\langle\hat{\mathbf{J}}\rangle$ may have arbitrary direction in 
space. We calculate the 
variances
\begin{equation}
\Delta{J}_{1,2}^2 = \langle\hat{J}_{1,2}^2\rangle -
\langle\hat{J}_{1,2}\rangle^2,
\label{1.2a1}
\end{equation}
 where
$\hat{J}_{1}$ and $\hat{J}_{2}$ are the components of 
$\hat{\mathbf J}$ along two mutually perpendicular directions in a
plane perpendicular to $\langle\hat{\mathbf{J}}\rangle$.
If these variances satisfy
\begin{equation}
\Delta{J}_{1}^2 = \Delta{J}_{2}^2 = j/2 = N/4,
\label{1.2new2} 
\end{equation} 
then the state $|\psi_j\rangle$ is called an atomic coherent state.
If 
\begin{equation}
\Delta{J}_{1}^2~~or,~~\Delta{J}_{2}^2~~ < j/2 = N/4,
\end{equation}
the state $|\psi_j\rangle$ is said to be an atomic squeezed state or
spin squeezed state \cite{Kitagawa}.
 This definition of spin squeezing is free from the
coordinate dependency and includes quantum correlation among the
atoms in the notion of squeezing. 
We can now define quantities
\begin{equation} 
Q_1 = \sqrt{\frac{2}{j}}\Delta{J}_1
\label{1.4new1}
\end{equation}
and 
\begin{equation}
Q_2 = \sqrt{\frac{2}{j}}\Delta{J}_2 
\label{1.5new2}
\end{equation}
such that if $Q_1$ and $Q_2$ are
equal to $1$, then $|\psi_j\rangle$ is called an atomic coherent state. If 
\begin{equation}
Q_1 ~~ or, ~~ Q_2~~ < 1,
\label{1.6newa1}
\end{equation}
the state $|\psi_j\rangle$ is said to be an atomic squeezed state \cite{Wineland}.

It is to be mentioned here that
\begin{equation}
Q_1~Q_2 \ge 1,
\label{1.6newa2}
\end{equation}
which is Heisenberg's uncertainty principle.

Now, normally, to perform the above calculations, we
rotate the coordinate system $\{x,y,z\}$ to 
$\{x^\prime,y^\prime,z^\prime\}$, such that the mean pseudo-spin
vector $\langle\hat{\mathbf{J}}\rangle$ points along the $z^\prime$
axis. We then calculate the variances 
\begin{equation}
\Delta{J}_{x^{\prime},y^{\prime}}^2 = 
\langle\hat{J}_{x^\prime,y^\prime}^2\rangle -
\langle\hat{J}_{x^\prime,y^\prime}\rangle^2
\label{1.6a3}
\end{equation}
and investigate the behavior of 
 \begin{equation} 
Q_x = \sqrt{\frac{2}{j}}\Delta{J}_{x^\prime}
\label{1.4}
\end{equation}
and 
\begin{equation}
Q_y = \sqrt{\frac{2}{j}}\Delta{J}_{y^\prime}. 
\label{1.5}
\end{equation}
If
\begin{equation}
Q_x = Q_y = 1,
\end{equation}
the corresponding quantum state of the system is a coherent state.
If 
\begin{equation}
Q_x ~~ or, ~~ Q_y~~ < 1,
\label{1.6a1}
\end{equation}
the quantum state is a spin squeezed state. Here, $Q_x$ and $Q_y$ satisfy
\begin{equation}
Q_x~Q_y \ge 1.
\label{1.6a2}
\end{equation}

A collective state vector $|\alpha\rangle$ for a system of two atoms is said to be quantum mechanically entangled  if 
$|\alpha\rangle$ cannot be expressed as a direct product
of the two individual atomic state vectors, i.e.,
\begin{equation}
|\alpha\rangle \ne |\alpha_1\rangle \otimes|\alpha_2\rangle,
\label{1.7}
\end{equation}
where $|\alpha_1\rangle$ and $|\alpha_2\rangle$ are the 
state vectors of the two individual atoms \cite{Nielsen}.

This paper is organized as follows.
In section II we formulate the necessary and sufficient condition for
the presence of quantum entanglement in arbitrary symmetric pure states
of two two-level atoms. We also construct a parameter, called quantum
entanglement parameter,
to quantify quantum entanglement in such systems. 
In section III we establish the relationship between quantum entanglement parameter and experimentally measurable quantities. That
is, we show how the quantum entanglement parameter can be measured
experimentally. 
  In section IV, we generalize these ideas in case of
systems containing $N$ number of two-level atoms.

\section{Quantum entanglement in a system of two two-level atoms }

We consider a system of two two-level atoms. 
To formulate the necessary and sufficient condition for the presence
of quantum entanglement in arbitrary symmetric pure states of this system, we first find out the quantum fluctuations of the composite system in terms of the components of
$\hat{\mathbf J}$, along two mutually orthogonal directions in a 
plane perpendicular to $\langle\hat{\mathbf J}\rangle$. 
 We then express these fluctuations as an algebraic sum of the quantum
fluctuations of the individual constituent atoms and their correlation terms. This helps us to isolate and study solely the quantum
correlation terms among the two atoms and
obtain the necessary and sufficient condition for the presence of quantum entanglement. From there we also construct a parameter to quantify quantum entanglement.

A normalized collective quantum state  $|\psi\rangle$ for this system
 can be expressed as
\begin{eqnarray}
|\psi\rangle &=& C_1 |j=1,m=1\rangle + C_2 |j=1,m=0\rangle
\nonumber\\
&+& C_3 |j=1,m=-1\rangle,
\label{2.1new1}
\end{eqnarray}
where $C_1$, $C_2$, and $C_3$ are constants satisfying
\begin{equation}
|C_1|^2 + |C_2|^2 + |C_3|^2 = 1.
\label{2.2new1}
\end{equation}
The state $|j=1,m=1\rangle$ corresponds to the case when
both the atoms are in their respective upper states. 
$|j=1,m=-1\rangle$
means both the atoms are in their lower states and 
$|j=1,m=0\rangle$ implies one atom in the upper and
 the other in its lower state.

In $\{m_1, m_2\}$ representation, if $|m_1, m_2\rangle$ is
the simultaneous eigenvector of $\hat{J}_{1_{z}}$ and
 $\hat{J}_{2_{z}}$ with eigenvalues $m_1$ and
$m_2$, respectively, then we can write
\begin{equation}
|j=1,m=1\rangle = |m_1=1/2, m_2=1/2\rangle,
\label{2.3new1}
\end{equation}
\begin{equation}
|j=1,m=-1\rangle = |m_1=-1/2, m_2=-1/2\rangle,
\label{2.4new1}
\end{equation}
and
\begin{eqnarray}
|j=1,m=0\rangle &=& \frac{1}{\sqrt{2}}
\Big[|m_1=1/2, m_2=-1/2\rangle \nonumber\\
&+& |m_1=-1/2, m_2=1/2\rangle\Big]
\label{2.5new1}
\end{eqnarray}
\cite{Sakurai}.
Thus, $|\psi\rangle$ in Eq. (\ref{2.1new1}) can be written as

\begin{eqnarray}
|\psi\rangle &=& C_1|1/2,1/2\rangle +
\frac{1}{\sqrt{2}}
C_2\Big[|1/2, -1/2\rangle\nonumber\\
&+& |-1/2, 1/2\rangle\Big] + C_3|-1/2, -1/2\rangle.
\label{2.6new1}
\end{eqnarray}
This state vector is symmetric under the exchange of two atoms.

   Since $|\psi\rangle$ is arbitrary, 
the quantities $\langle\psi|\hat{J}_x|\psi\rangle$,
$\langle\psi|\hat{J}_y|\psi\rangle$, and 
$\langle\psi|\hat{J}_z|\psi\rangle$ have arbitrary values and, hence,
$\langle\hat{\mathbf{J}}\rangle = 
\langle\psi|\hat{\mathbf{J}}|\psi\rangle$
 has arbitrary direction.  We now perform a rotation 
of the coordinate system from $\{x,y,z\}$ to
$\{x^\prime, y^\prime, z^\prime\}$ such that the vector
$\langle\hat{\mathbf{J}}\rangle$ points along the $z^\prime$ axis.
In doing so, we assume that the vector 
$\langle\hat{\mathbf{J}}\rangle$ was in the first octant of the
coordinate system $\{x,y,z\}$.  After the rotation, the components 
$\{\hat{J}_{x^\prime}, \hat{J}_{y^\prime}, \hat{J}_{z^\prime}\}$ in the rotated frame $\{x^\prime, y^\prime, z^\prime\}$
 are related to $\{\hat{J}_x, \hat{J}_y, \hat{J}_z\}$
 in the unrotated frame $\{x, y, z\}$ as  
\begin{eqnarray}
\hat{J}_{x^\prime} &=& \hat{J}_x\cos\theta\cos\phi + \hat{J}_y
\cos\theta\sin\phi - \hat{J}_z\sin\theta,\label{2.3}\\
\hat{J}_{y^\prime} &=& -\hat{J}_x\sin\phi + \hat{J}_y\cos\phi,
\label{2.4}\\
\hat{J}_{z^\prime} &=& \hat{J}_x\sin\theta\cos\phi + \hat{J}_y
\sin\theta\sin\phi + \hat{J}_z\cos\theta,\label{2.5}
\end{eqnarray}
where
\begin{eqnarray}
\cos\theta &=& \frac{\langle\hat{J}_z\rangle}
{|\langle\hat{\mathbf{J}}\rangle|},\label{2.6}\\
\cos\phi &=& \frac{\langle\hat{J}_x\rangle}{\sqrt{\langle\hat{J}_x\rangle^2 + \langle\hat{J}_y\rangle^2}}.
\label{2.7}
\end{eqnarray}
We can check using Eqs. $(\ref{2.3})$, $(\ref{2.6})$, and 
$(\ref{2.7})$,
that for the arbitrary state $|\psi\rangle$ we have
\begin{eqnarray}
\langle\hat{J}_{x^\prime}\rangle &=& \langle\hat{J}_x\rangle
\cos\theta\cos\phi + \langle\hat{J}_y\rangle
\cos\theta\sin\phi\nonumber\\
 &&- \langle\hat{J}_z\rangle\sin\theta\label{2.8}\\
&=& \frac{1}{|\langle\hat{\mathbf J}\rangle|\sqrt{\langle\hat{J}_x
\rangle^2 + \langle\hat{J}_y\rangle^2}}\bigg[\langle\hat{J}_x
\rangle^2\langle\hat{J}_z\rangle + \langle\hat{J}_y
\rangle^2\langle\hat{J}_z\rangle\nonumber\\
&& - \langle\hat{J}_z\rangle
\Big(\langle\hat{J}_x\rangle^2 + \langle\hat{J}_y\rangle^2\Big)\bigg]
\nonumber\\
&=& 0. 
\label{2.9}
\end{eqnarray}
Similarly, using Eqs. $(\ref{2.4})$, $(\ref{2.5})$, $(\ref{2.6})$, and $(\ref{2.7})$, we have,
\begin{eqnarray}
\langle\hat{J}_{y^\prime}\rangle &=& 0,\label{2.9new1}\\
\langle\hat{J}_{z^\prime}\rangle &=& |\langle\hat{\mathbf J}\rangle|.
\label{2.10}
\end{eqnarray}
Thus, the mean pseudo-spin vector is now along the $z^\prime$ axis.

We now calculate the quantum fluctuations in the components of
$\hat{\mathbf J}$ along two mutually orthogonal directions, in a plane
perpendicular to $\langle\hat{\mathbf J}\rangle$. For simplicity,
we take the above-mentioned two orthogonal directions along the 
$x^\prime$ and $y^\prime$ axes, respectively. Therefore, we calculate the quantum
fluctuations
$\Delta{J}_{x^\prime}^2$ and 
$\Delta{J}_{y^\prime}^2$.

Now, we have already shown in Eqs. $(\ref{2.9})$ and 
$(\ref{2.9new1})$ that we have here $\langle\hat{J}_{x^\prime}\rangle = \langle\hat{J}_{y^\prime}\rangle = 0$. Therefore, according to Eqs. $(\ref{1.6a3})$, $(\ref{2.3})$, and $(\ref{2.4})$, we obtain for the quantum state 
$|\psi\rangle$,
\begin{eqnarray}
\Delta{J_{x^\prime}^2} &=& \langle\hat{J}_{x^\prime}^2\rangle = 
\langle\hat{J}_x^2\rangle
\cos^2\theta\cos^2\phi + 
\langle\hat{J}_y^2\rangle\cos^2\theta\sin^2\phi\nonumber\\
 &+& \langle\hat{J}_z^2
\rangle\sin^2\theta\nonumber\\
&+& \frac{1}{2}\langle\hat{J}_x\hat{J}_y + \hat{J}_y\hat{J}_x\rangle\cos^2\theta\sin2\phi\nonumber\\
&-& \frac{1}{2}\langle\hat{J}_x\hat{J}_z + \hat{J}_z\hat{J}_x\rangle\sin2\theta
\cos\phi\nonumber\\
 &-& \frac{1}{2}\langle\hat{J}_y\hat{J}_z + \hat{J}_z\hat{J}_y\rangle
\sin2\theta\sin\phi
\label{2.12}
\end{eqnarray}
and
\begin{eqnarray}
\Delta{J_{y^\prime}^2} &=& \langle\hat{J}_{y^\prime}^2\rangle =  \langle\hat{J}_x^2\rangle
\sin^2\phi + 
\langle\hat{J}_y^2\rangle\cos^2\phi\nonumber\\
 &-& \frac{1}{2}\langle\hat{J}_x\hat{J}_y
+ \hat{J}_y\hat{J}_x\rangle\sin2\phi,
\label{2.12new1}
\end{eqnarray}
respectively.

It is to be mentioned here that we are not using the forms of
$\cos\theta$, $\cos\phi$, etc., as given in Eqs. $(\ref{2.6})$ and
$(\ref{2.7})$, respectively, as to keep the mathematical expressions neat. At the
end of the calculation, we use the above-mentioned equations to ensure that we are calculating the fluctuations in a plane perpendicular to
$\langle\hat{\mathbf J}\rangle$.

 We now express these fluctuations as an algebraic sum of the
quantum fluctuations of the individual constituent atoms and their
correlation terms.
 From Eq. $(\ref{1.1a})$, we have for a system of two two-level
atoms,
\begin{eqnarray}
\hat{J}_x &=& \hat{J}_{1_x} + \hat{J}_{2_x},\label{2.16}\\
\hat{J}_y &=& \hat{J}_{1_y} + \hat{J}_{2_y},\label{2.16a1}\\
\hat{J}_z &=& \hat{J}_{1_z} + \hat{J}_{2_z}.
\label{2.16a2}
\end{eqnarray} 
Therefore, we have
\begin{eqnarray}
\langle\hat{J}_{x,y,z}^2\rangle &=& \langle\hat{J}_{1_{x,y,z}}^2
\rangle + 
\langle\hat{J}_{2_{x,y,z}}^2\rangle\nonumber\\
&+& \langle\hat{J}_{1_{x,y,z}}\hat{J}_{2_{x,y,z}}
+ \hat{J}_{2_{x,y,z}}\hat{J}_{1_{x,y,z}}\rangle.
\label{2.17}
\end{eqnarray}

Now, using Eqs. $(\ref{2.16})$ and $(\ref{2.16a1})$, we have
\begin{eqnarray}
\langle\hat{J}_x\hat{J}_y + \hat{J}_y\hat{J}_x\rangle &=&
\langle\hat{J}_{1_x}\hat{J}_{1_y} + \hat{J}_{1_y}
\hat{J}_{1_x}\rangle + \langle\hat{J}_{2_x}\hat{J}_{2_y} + \hat{J}_{2_y}
\hat{J}_{2_x}\rangle\nonumber\\
&+&  \langle\hat{J}_{1_x}\hat{J}_{2_y} + 
\hat{J}_{2_y}\hat{J}_{1_x}\rangle +
\langle\hat{J}_{1_y}\hat{J}_{2_x} + \hat{J}_{2_x}\hat{J}_{1_y}\rangle.
\nonumber\\
\label{2.18}
\end{eqnarray}

Similarly, using Eqs. $(\ref{2.16})$, $(\ref{2.16a1})$,
 and $(\ref{2.16a2})$, we have
\begin{eqnarray}
\langle\hat{J}_x\hat{J}_z + \hat{J}_z\hat{J}_x\rangle &=&
\langle\hat{J}_{1_x}\hat{J}_{1_z} + \hat{J}_{1_z}
\hat{J}_{1_x}\rangle + \langle\hat{J}_{2_x}\hat{J}_{2_z} + \hat{J}_{2_z}
\hat{J}_{2_x}\rangle\nonumber\\
&+& \langle\hat{J}_{1_x}\hat{J}_{2_z} + \hat{J}_{2_z}\hat{J}_{1_x}
\rangle + 
\langle\hat{J}_{1_z}\hat{J}_{2_x} + \hat{J}_{2_x}\hat{J}_{1_z}\rangle
\nonumber\\
\label{2.18a1}
\end{eqnarray}
and
\begin{eqnarray}
\langle\hat{J}_y\hat{J}_z &+& \hat{J}_z\hat{J}_y\rangle =
\langle\hat{J}_{1_y}\hat{J}_{1_z} + \hat{J}_{1_z}
\hat{J}_{1_y}\rangle + \langle\hat{J}_{2_y}\hat{J}_{2_z} + \hat{J}_{2_z}
\hat{J}_{2_y}\rangle\nonumber\\
&+& \langle\hat{J}_{1_y}\hat{J}_{2_z} + \hat{J}_{2_z}\hat{J}_{1_y}
\rangle + \langle\hat{J}_{1_z}\hat{J}_{2_y} + 
\hat{J}_{2_y}\hat{J}_{1_z}\rangle.\nonumber\\
\label{2.18a2}
\end{eqnarray}

It is to be noted here that though the operators of atom 1 commute
with those of atom 2, we are not taking advantage of that as to keep
the expressions symmetric with respect to the indices 1 and 2.

Using Eqs. $(\ref{2.17})$ to $(\ref{2.18a2})$ in Eq. $(\ref{2.12})$
and $(\ref{2.12new1})$, we get
\begin{eqnarray}
\Delta{J_{x^\prime}^2} &=& \sum_{i=1}^{2}\bigg[\langle\hat{J}_{i_x}^2
\rangle\cos^2\theta\cos^2\phi + \langle\hat{J}_{i_y}^2
\rangle\cos^2\theta\sin^2\phi\nonumber\\
&+& \langle\hat{J}_{i_z}^2
\rangle\sin^2\theta\bigg]
+ \sum_{i=1}^{2}\sum_{{}^{l=1}_{l\ne i}}^{2}\bigg[\langle\hat{J}_{i_x}
\hat{J}_{l_x}\rangle\cos^2\theta\cos^2\phi\nonumber\\
 &+& \langle\hat{J}_{i_y}
\hat{J}_{l_y}\rangle\cos^2\theta\sin^2\phi + \langle\hat{J}_{i_z}
\hat{J}_{l_z}\rangle\sin^2\theta\bigg]\nonumber\\
&+& \frac{1}{2}\sum_{i=1}^{2}\sum_{l=1}^{2}\bigg[\langle\hat{J}_{i_x}
\hat{J}_{l_y} + \hat{J}_{l_y}
\hat{J}_{i_x}\rangle\cos^2\theta\sin2\phi\nonumber\\
&-& \langle\hat{J}_{i_x}
\hat{J}_{l_z} + \hat{J}_{l_z}
\hat{J}_{i_x}\rangle\sin2\theta\cos\phi\nonumber\\
&-& \langle\hat{J}_{i_y}
\hat{J}_{l_z} + \hat{J}_{l_z}
\hat{J}_{i_y}\rangle\sin2\theta\sin\phi\bigg]
\label{2.18a3}
\end{eqnarray}
and
\begin{eqnarray}
\Delta{J_{y^\prime}^2} &=& \sum_{i=1}^{2}\bigg[\langle\hat{J}_{i_x}^2
\rangle\sin^2\phi + \langle\hat{J}_{i_y}^2\rangle\cos^2\phi\bigg]
\nonumber\\
&+& \sum_{i=1}^{2}\sum_{{}^{l=1}_{l\ne i}}^{2}\bigg[
\langle\hat{J}_{i_x}
\hat{J}_{l_x}\rangle\sin^2\phi + \langle\hat{J}_{i_y}\hat{J}_{l_y}
\rangle\cos^2\phi\bigg]\nonumber\\
&-& \frac{1}{2}\sum_{i=1}^{2}\sum_{l=1}^{2}\langle\hat{J}_{i_x}
\hat{J}_{l_y} + \hat{J}_{l_y}\hat{J}_{i_x}\rangle\sin2\phi.
\label{2.18a4}
\end{eqnarray}

 Now, since $|\psi\rangle$ is symmetric under the exchange of two atoms and both the atoms have been treated on equal footing in
the state $|\psi\rangle$, we have
\begin{eqnarray}
\langle\hat{J}_{1_{x}}\rangle &=& \langle\hat{J}_{2_{x}}
\rangle,\label{2.25new7}\\
\langle\hat{J}_{1_{y}}\rangle &=& \langle\hat{J}_{2_{y}}
\rangle, \label{2.25new8}\\
\langle\hat{J}_{1_{z}}\rangle &=& \langle\hat{J}_{2_{z}}
\rangle, \label{2.25new9}\\
\langle\hat{J}_{1_x}\hat{J}_{2_y}\rangle &=&
\langle\hat{J}_{1_y}\hat{J}_{2_x}\rangle,
\label{2.19a3}\\
\langle\hat{J}_{1_x}\hat{J}_{2_z}\rangle &=& 
\langle\hat{J}_{1_z}\hat{J}_{2_x}\rangle,\label{2.19a4}\\
\langle\hat{J}_{1_y}\hat{J}_{2_z}\rangle &=& 
\langle\hat{J}_{1_z}\hat{J}_{2_y}\rangle,\label{2.19a5}\\
\langle\hat{J}_{1_x}\hat{J}_{1_y} + \hat{J}_{1_y}
\hat{J}_{1_x}\rangle &=& \langle\hat{J}_{2_x}\hat{J}_{2_y} + 
\hat{J}_{2_y}\hat{J}_{2_x}\rangle,\label{2.19a6}\\
\langle\hat{J}_{1_x}\hat{J}_{1_z} + \hat{J}_{1_z}
\hat{J}_{1_x}\rangle &=& \langle\hat{J}_{2_x}\hat{J}_{2_z} + 
\hat{J}_{2_z}\hat{J}_{2_x}\rangle,\label{2.19a7}\\
\langle\hat{J}_{1_y}\hat{J}_{1_z} + \hat{J}_{1_z}
\hat{J}_{1_y}\rangle &=& \langle\hat{J}_{2_y}\hat{J}_{2_z} + 
\hat{J}_{2_z}\hat{J}_{2_y}\rangle.\label{2.19a8}
\end{eqnarray}

Therefore, using Eqs. $(\ref{2.25new7})$, $(\ref{2.25new8})$, and
$(\ref{2.25new9})$, we can reduce $\cos\theta$ and $\cos\phi$ given
in Eqs. $(\ref{2.6})$ and $(\ref{2.7})$, respectively, as
\begin{eqnarray}
\cos\theta &=& \frac{\langle\hat{J}_{1_z}\rangle}
{|\langle\hat{\mathbf{J}}_1\rangle|},\label{2.19new8}\\
\cos\phi &=& \frac{\langle\hat{J}_{1_x}\rangle}
{\sqrt{\langle\hat{J}_{1_x}\rangle^2 + \langle\hat{J}_{1_y}
\rangle^2}},
\label{2.19new9}
\end{eqnarray}
where
\begin{equation}
|\langle\hat{\mathbf{J}}_1\rangle| = \sqrt{\langle\hat{J}_{1_x}\rangle^2 + \langle\hat{J}_{1_y}\rangle^2 + \langle\hat{J}_{1_z}
\rangle^2}.
\label{2.19newa1}
\end{equation}
Using Eqs. $(\ref{2.3})$, $(\ref{2.16})$, $(\ref{2.19new8})$, and
$(\ref{2.19new9})$, it can be shown that
\begin{eqnarray}
\langle\hat{J}_{i_{x^\prime}}\rangle &=& \langle\hat{J}_{i_x}\rangle
\cos\theta\cos\phi + \langle\hat{J}_{i_y}\rangle
\cos\theta\sin\phi - \langle\hat{J}_{i_z}\rangle
\sin\theta\nonumber\\
 &=& 0.
\label{2.19new10}
\end{eqnarray} 
Therefore, from Eqs. $(\ref{1.6a3})$ and $(\ref{2.3})$, we have
\begin{eqnarray}
\Delta{J_{i_{x^\prime}}^2} &=& \langle\hat{J}_{i_{x^\prime}}^2\rangle = \langle\hat{J}_{i_x}^2\rangle\cos^2\theta\cos^2\phi\nonumber\\
&+& \langle\hat{J}_{i_y}^2\rangle\cos^2\theta\sin^2\phi
+ \langle\hat{J}_{i_z}^2\rangle\sin^2\theta\nonumber\\
&+& \frac{1}{2}\langle\hat{J}_{i_x}\hat{J}_{i_y} + \hat{J}_{i_y}
\hat{J}_{i_x}\rangle\cos^2\theta\sin2\phi\nonumber\\
&-& \frac{1}{2}\langle\hat{J}_{i_x}\hat{J}_{i_z} + \hat{J}_{i_z}
\hat{J}_{i_x}\rangle\sin2\theta\cos\phi\nonumber\\
 &-& \frac{1}{2}\langle\hat{J}_{i_y}\hat{J}_{i_z} + \hat{J}_{i_z}
\hat{J}_{i_y}\rangle\sin2\theta\sin\phi,
\label{2.19new10a1}
\end{eqnarray}
where, $i = 1, 2$.
Using the above equation we can write Eq. $(\ref{2.18a3})$ as
\begin{eqnarray}
\Delta{J_{x^\prime}^2} &=&
\sum_{i=1}^{2} \Delta{J_{i_{x^\prime}}^2} + \sum_{i=1}^{2}
\sum_{{}^{l=1}_{l\ne i}}^{2}\bigg[\langle\hat{J}_{i_{x}}
\hat{J}_{l_{x}}\rangle\cos^2\theta\cos^2\phi\nonumber\\
&+& \langle\hat{J}_{i_{y}}
\hat{J}_{l_{y}}\rangle\cos^2\theta\sin^2\phi
+ \langle\hat{J}_{i_{z}}
\hat{J}_{l_{z}}\rangle\sin^2\theta\nonumber\\
&+& \langle\hat{J}_{i_{x}}
\hat{J}_{l_y}\rangle\cos^2\theta\sin2\phi
 -\langle\hat{J}_{i_{x}}
\hat{J}_{l_{z}}\rangle\sin2\theta\cos\phi \nonumber\\
&-& \langle\hat{J}_{i_{y}}
\hat{J}_{l_{z}}\rangle\sin2\theta\sin\phi\bigg].
\label{2.19newm1}  
\end{eqnarray}

Similarly, it can be shown that
\begin{eqnarray}
\Delta{J}_{i_{y^\prime}}^2 &=& \langle\hat{J}_{i_{y^\prime}}^2\rangle = \langle\hat{J}_{i_x}^2\rangle\sin^2\phi
+\langle\hat{J}_{i_y}^2\rangle\cos^2\phi\nonumber\\
&-& \frac{1}{2}\langle\hat{J}_{i_x}\hat{J}_{i_y} + \hat{J}_{i_y}\hat{J}_{i_x}\rangle\sin2\phi
\label{2.19new10a4}
\end{eqnarray}
and, hence, we have
\begin{eqnarray}
\Delta{J_{y^\prime}^2} &=&
\sum_{i=1}^{2} \Delta{J_{i_{y^\prime}}^2} + \sum_{i=1}^{2}
\sum_{{}^{l=1}_{l\ne i}}^{2}\bigg[ \langle\hat{J}_{i_{x}}
\hat{J}_{l_{x}}\rangle\sin^2\phi\nonumber\\
&+& \langle\hat{J}_{i_{y}}
\hat{J}_{l_{y}}\rangle\cos^2\phi
-\langle\hat{J}_{i_{x}}\hat{J}_{l_{y}}\rangle\sin2\phi\bigg].
\label{2.19newm2}
\end{eqnarray}

We now take advantage of the fact that the operators of atom 1 commute with those of atom 2 and obtain, using Eqs. $(\ref{2.19a6})$ to
$(\ref{2.19a8})$ in Eq. $(\ref{2.19newm1})$, 

\begin{eqnarray}
\Delta{J}_{x^\prime}^2 &=& \Delta{J}_{1_{x^\prime}}^2 + 
\Delta{J}_{2_{x^\prime}}^2 +
2\langle\hat{J}_{1_x}\hat{J}_{2_x}\rangle
\cos^2\theta\cos^2\phi\nonumber\\ 
&+& 2\langle\hat{J}_{1_y}
\hat{J}_{2_y}\rangle\cos^2\theta\sin^2\phi
+ 2\langle\hat{J}_{1_z}
\hat{J}_{2_z}\rangle\sin^2\theta\nonumber\\
&+& 2\langle\hat{J}_{1_x}
\hat{J}_{2_y}\rangle\cos^2\theta\sin2\phi
- 2 \langle\hat{J}_{1_x}\hat{J}_{2_z}\rangle
\sin2\theta\cos\phi\nonumber\\
&-& 2 \langle\hat{J}_{1_y}\hat{J}_{2_z}\rangle
\sin2\theta\sin\phi.\label{2.20new1} 
 \end{eqnarray}

Thus, the quantum fluctuation $\Delta{J}_{x^\prime}^2$ of a
composite system of two two-level atoms is equal to the
sum of the fluctuations 
$\Delta{J}_{1_{x^\prime}}^2$ and $\Delta{J}_{2_{x^\prime}}^2$
of the individual constituent atoms and the correlation terms 
$\langle\hat{J}_{1_{x,y,z}}\hat{J}_{2_{x,y,z}}\rangle$,
$\langle\hat{J}_{1_x}\hat{J}_{2_y}\rangle$,
$\langle\hat{J}_{1_x}\hat{J}_{2_z}\rangle$, and
$\langle\hat{J}_{1_y}\hat{J}_{2_z}\rangle$, which depend upon the correlation among the two atoms.

In a similar fashion, we have
\begin{eqnarray}
\Delta{J}_{y^\prime}^2 &=& \Delta{J}_{1_{y^\prime}}^2 + 
\Delta{J}_{2_{y^\prime}}^2 + 
2\langle\hat{J}_{1_y}\hat{J}_{2_y}\rangle\cos^2\phi\nonumber\\
 &+& 2\langle\hat{J}_{1_x}\hat{J}_{2_x}\rangle\sin^2\phi
 - 2 \langle\hat{J}_{1_x}\hat{J}_{2_y}\rangle
\sin2\phi,\label{2.20new2}
\end{eqnarray}
where the last three terms represent the correlation among the two atoms.

Thus, we can see from Eqs. $(\ref{2.20new1})$ and $(\ref{2.20new2})$
 that, by expressing the quantum fluctuations 
$\Delta{J}_{x^\prime}^2$ and $\Delta{J}_{y^\prime}^2$ of the
 composite system of two two-level atoms in the above way, we can separate out the correlation terms among the two atoms from their
 intrinsic quantum fluctuations. This helps to visualize and study solely the quantum correlations existing among the two atoms.

We now calculate $\Delta{J}_{1_{x^\prime,y^\prime}}^2$ and 
$\Delta{J}_{2_{x^\prime,y^\prime}}^2$ for the
state $|\psi\rangle$.
 Using the expression of $|\psi\rangle$ given in 
Eq. $(\ref{2.6new1})$, we get
\begin{eqnarray}
&& \langle\hat{J}_{i_{x}}^2\rangle = 
\langle\hat{J}_{i_{y}}^2\rangle =
 \langle\hat{J}_{i_{z}}^2\rangle = \frac{1}{4}
\label{2.25new3}
\end{eqnarray}
and
\begin{eqnarray}
&&\langle\hat{J}_{i_{x}}\hat{J}_{i_{y}} + \hat{J}_{i_{y}}
\hat{J}_{i_{x}}\rangle = 
\langle\hat{J}_{i_{x}}\hat{J}_{i_{z}} + \hat{J}_{i_{z}}
\hat{J}_{i_{x}}\rangle\nonumber\\
&&=\langle\hat{J}_{i_{y}}\hat{J}_{i_{z}} + \hat{J}_{i_{z}}
\hat{J}_{i_{y}}\rangle = 0,
\label{2.25new4}
\end{eqnarray}
where $i = 1, 2$.
Therefore, using these equations and also Eqs. $(\ref{2.19new8})$ and $(\ref{2.19new9})$ in
Eq. $(\ref{2.19new10a1})$, we get
\begin{equation}
\Delta J_{i_{x^\prime}}^2 = \Delta J_{i_{y^\prime}}^2 = 
\frac{1}{4}.
\label{2.25new5}
\end{equation}
Now, using the expressions of $\cos\theta$ and $\cos\phi$ given in Eqs. $(\ref{2.19new8})$ and $(\ref{2.19new9})$, respectively, and also
using Eq.
$(\ref{2.25new5})$ in Eq. $(\ref{2.20new1})$,
we obtain
\begin{eqnarray}
&&\Delta{J}_{x^\prime}^2 = \frac{1}{4} + \frac{1}{4} + 
\frac{2\langle\hat{J}_{1_z}\rangle^2}{|\langle\hat{\mathbf J}_{1}\rangle|^2\Big(\langle\hat{J}_{1_x}
\rangle^2 + \langle\hat{J}_{1_y}\rangle^2\Big)}\nonumber\\
&\times& \bigg[\langle\hat{J}_{1_x}\hat{J}_{2_x}\rangle
\langle\hat{J}_{1_x}\rangle^2 + 2\langle\hat{J}_{1_x}\hat{J}_{2_y}
\rangle\langle\hat{J}_{1_x}\rangle\langle\hat{J}_{1_y}\rangle
\nonumber\\ 
&+& 
\langle\hat{J}_{1_y}\hat{J}_{2_y}\rangle\langle\hat{J}_{1_y}\rangle^2
\bigg] + \frac{2\langle\hat{J}_{1_z}\hat{J}_{2_z}\rangle}
{|\langle\hat{\mathbf J}_{1}\rangle|^2}\bigg[\langle\hat{J}_{1_x}
\rangle^2 + \langle\hat{J}_{1_y}\rangle^2\bigg]\nonumber\\
&-& \frac{4\langle\hat{J}_{1_z}\rangle}{|\langle\hat{\mathbf J}_{1}\rangle|^2}\bigg[\langle\hat{J}_{1_x}\hat{J}_{2_z}\rangle
\langle\hat{J}_{1_x}\rangle + \langle\hat{J}_{1_y}\hat{J}_{2_z}\rangle
\langle\hat{J}_{1_y}\rangle \bigg]\label{2.20}\\
&=& \frac{1}{2} + corrx, 
\label{2.25new19}
\end{eqnarray}
where $corrx$ is the sum of last seven terms in 
Eq. $(\ref{2.20})$.
It represents the quantum correlation existing among the two atoms.

Similarly, using Eqs. $(\ref{2.19new8})$, $(\ref{2.19new9})$, and 
$(\ref{2.25new5})$ in Eq. $(\ref{2.20new2})$, we get
\begin{eqnarray}
\Delta{J}_{y^\prime}^2 &=& \frac{1}{4} + \frac{1}{4} + 
\frac{2}{\Big(\langle\hat{J}_{1_x}
\rangle^2 + \langle\hat{J}_{1_y}\rangle^2\Big)}
\bigg[\langle\hat{J}_{1_x}\hat{J}_{2_x}\rangle
\langle\hat{J}_{1_y}\rangle^2 \nonumber\\
&+& \langle\hat{J}_{1_y}\hat{J}_{2_y}
\rangle\langle\hat{J}_{1_x}\rangle^2 - 2\langle\hat{J}_{1_x}
\hat{J}_{2_y}\rangle\langle\hat{J}_{1_x}\rangle\langle\hat{J}_{1_y}
\rangle\bigg]\label{2.21}\\ 
&=& \frac{1}{2} + corry,
\label{2.25new20}
\end{eqnarray} 
where $corry$ represents the correlation among the two atoms.

We now see what happens to the correlation terms $corrx$ and $corry$
for an unentangled state.
As mentioned in the earlier section, an unentangled state 
$|\psi\rangle$ of the composite system of two atoms can be expressed
as a direct product of the individual atomic state vectors of the two 
constituent atoms as
\begin{equation}
|\psi\rangle = |\psi_1\rangle \otimes |\psi_2\rangle,
\label{2.22}
\end{equation} 
where $|\psi_1\rangle$ and $|\psi_2\rangle$ are the atomic state
vectors corresponding to the two constituent atoms. 
Now, it is easy to see that for these kinds of states the following
conditions are satisfied.
 
\begin{eqnarray}
\langle\hat{J}_{1_x}\hat{J}_{2_x}\rangle &=& 
\langle\hat{J}_{1_x}\rangle\langle\hat{J}_{2_x}\rangle,~~
\langle\hat{J}_{1_y}\hat{J}_{2_y}\rangle =
\langle\hat{J}_{1_y}\rangle\langle\hat{J}_{2_y}\rangle,
\nonumber\\
\langle\hat{J}_{1_z}\hat{J}_{2_z}\rangle &=& 
\langle\hat{J}_{1_z}\rangle\langle\hat{J}_{2_z}\rangle,~~
\langle\hat{J}_{1_x}\hat{J}_{2_y}\rangle = 
\langle\hat{J}_{1_x}\rangle\langle\hat{J}_{2_y}\rangle,\nonumber\\
\langle\hat{J}_{1_x}\hat{J}_{2_z}\rangle &=& 
\langle\hat{J}_{1_x}\rangle\langle\hat{J}_{2_z}\rangle,~~
\langle\hat{J}_{1_y}\hat{J}_{2_z}\rangle = 
\langle\hat{J}_{1_y}\rangle\langle\hat{J}_{2_z}\rangle.\nonumber\\
\label{2.23}
\end{eqnarray}
Hence, using the above equations and also Eqs. $(\ref{2.25new7})$,
$(\ref{2.25new8})$, and $(\ref{2.25new9})$ in the expressions of 
$corrx$ and
$corry$ as given in Eqs. $(\ref{2.20})$ and $(\ref{2.21})$,
respectively, we get
\begin{equation}
corrx = corry = 0.
\label{2.24}
\end{equation} 
Thus, for an unentangled state $corrx$ and $corry$ are zero, and we
have
\begin{eqnarray}
\Delta{J}_{x^\prime,y^\prime}^2\bigg|_{un-ent} = \Delta{J}_{1_{x^\prime,y^\prime}}^2 + \Delta{J}_{2_{x^\prime,y^\prime}}^2
= \frac{1}{2}.
\label{2.25}
\end{eqnarray}
That is, the quantum fluctuations of the composite state is just the
algebraic sum of the corresponding fluctuations of the individual constituent atoms.

The terms $corrx$ and $corry$ are non zero when the atomic state vector is entangled. We can give a physical interpretation of $corrx$ and $corry$ in the following way. If 
$\Delta{J}_{x^\prime}^2$ and $\Delta{J}_{y^\prime}^2$ are
 the fluctuations  of an entangled state, then using 
Eqs. $(\ref{2.25new19})$, $(\ref{2.25new20})$, and $(\ref{2.25})$, we can write
\begin{eqnarray}
corrx &=& \Delta{J}_{x^\prime}^2 - \frac{1}{2} \label{2.26new1}\\
&=& \Delta{J}_{x^\prime}^2 - \Delta{J}_{x^\prime}^2
\bigg|_{un-ent},\label{2.26}\\
corry &=& \Delta{J}_{y^\prime}^2 - \frac{1}{2}\label{2.27new1}\\
&=& \Delta{J}_{y^\prime}^2 - \Delta{J}_{y^\prime}^2\bigg|_{un-ent}.
\label{2.27}
\end{eqnarray}

Thus, $corrx$ and $corry$ are the measures of the deviations
of the quantum fluctuations of an entangled state from those of an
unentangled one. Whenever $corrx$ and $corry$ for a quantum state of a composite system are non zero, we can
conclude that the corresponding quantum state is an entangled state.
 We can construct a parameter out of $corrx$ and $corry$ for the
detection and quantification of quantum entanglement.  Since $corrx$ and $corry$ may have
 opposite signs, and also to treat both of them on equal footing, 
we construct a parameter $S$ as,
\begin{equation}
S = \frac{1}{2}\Big({corrx}^2 + {corry}^2\Big),
\label{2.28}
\end{equation}
such that the non zero value of $S$ implies the presence of quantum entanglement in the corresponding system. We call $S$ the quantum
entanglement parameter.  $S$ is the mean squared
deviation of the quantum fluctuations in the two quadratures 
($x^\prime$ and $y^\prime$ ) of a quantum mechanically
entangled state from those of an unentangled one.

 Thus, whenever we have 
\begin{equation}
S = 0,
\label{2.29}
\end{equation}
 the corresponding quantum state is unentangled,
and whenever we have
\begin{equation}
S > 0,
\label{2.30}
\end{equation}
the corresponding quantum state is entangled.   
The condition $S > 0$ is the necessary and sufficient condition for 
the presence of quantum entanglement. We can prove it in this way.
We know that whenever a quantum state for a composite system of two
two-level atoms is entangled, the corresponding quantum state vector
cannot be written as a direct product of the individual atomic
 state vectors. In that case, the conditions in Eq. $(\ref{2.23})$
are not satisfied and hence $corrx$ and $corry$ are non-zero, 
implying that $S > 0$. This shows that the condition $S > 0$ forms
the necessary condition for the presence of quantum entanglement.
We now prove that the condition is sufficient also for the
 presence of entanglement in this way. Whenever $S > 0$, either
$corrx$ or $corry$ or both of them are non-zero. This means that
all the conditions in Eq. $(\ref{2.23})$ are not satisfied,
implying that the corresponding quantum state vector cannot be 
expressed as a direct product of the individual atomic state
 vectors, and, hence, the quantum state is entangled. Thus, we have proved that the condition $S > 0$ forms the necessary and
sufficient condition for the presence of quantum entanglement.

We can see from Eqs. $(\ref{2.25new19})$ that,
if $corrx < 0$, then
\begin{equation}
\Delta{J}_{x^\prime}^2 < \frac{1}{2},
\label{2.31}
\end{equation}
and, hence, the quantum state $|\psi\rangle$ is a spin squeezed state having squeezing in the $x^\prime$ quadrature.  We see from 
Eq. $(\ref{2.25new20})$  that, at the same time we should have 
$corry > 0$ (no squeezing in the $y^\prime$ quadrature), so that
\begin{equation}
\Delta{J}_{y^\prime}^2 > \frac{1}{2},
\label{2.32}
\end{equation}
and Heisenberg's uncertainty principle [ Eq. $(\ref{1.6a2})$ ] is restored. Similarly, when $corry < 0$, there is squeezing in the 
$y^\prime$ quadrature and no squeezing in the $x^\prime$ one.

Now, we see from the above discussion that when we have spin squeezing in a quantum state, either $corrx$ or $corry$ is less than
zero and hence $S > 0$, implying the presence of quantum entanglement. Thus, whenever there is spin squeezing there is quantum 
entanglement. But the reverse is not true.
It may happen that some quantum state does not show
spin squeezing at all, that is, $corrx$ and $corry$ are never less than zero and instead they are always greater than zero. In that case also we have $S > 0$, implying the presence of quantum entanglement. This shows that a quantum state which is not spin squeezed may show quantum entanglement.

For the purpose of quantification of entanglement, we notice that
since $S$ is the mean squared deviation of the quantum fluctuations in
the two quadratures ($x^\prime$ and $y^\prime$) of an entangled state
from the corresponding fluctuations of an unentangled one, we can take $S$ itself to be
proportional to the amount of quantum entanglement in a system. As 
$S$ for a system increases, the entanglement of the system also
increases. As $S$ decreases, the entanglement also decreases. Thus,
the value of $S$ itself can be a measure of entanglement in a 
system. The question now arises about how to measure $S$. In the
next section, we  establish a connection between $S$ and experimentally
measurable quantities.  

\section{Relationship between the quantum entanglement 
parameter $S$ and the experimentally measurable quantities}

In this section, we show how we can measure the qunatum entanglement
parameter $S$. We rewrite Eqs. $(\ref{2.25new19})$ and 
$(\ref{2.25new20})$ as
\begin{eqnarray}
corrx &=& \Delta{J}_{x^\prime}^2 - \frac{1}{2}, \label{3.1}\\
corry &=& \Delta{J}_{y^\prime}^2 - \frac{1}{2}. \label{3.2}
\end{eqnarray}
Therefore,
\begin{eqnarray}
corrx^2 &=& \Delta{J}_{x^\prime}^4 - \Delta{J}_{x^\prime}^2 
+ \frac{1}{4}\label{3.3}\\
corry^2 &=& \Delta{J}_{y^\prime}^4 - \Delta{J}_{y^\prime}^2 
+ \frac{1}{4}\label{3.4}
\end{eqnarray}
Hence,
\begin{eqnarray}
S &=& \frac{1}{2}\Big[ \Delta{J}_{x^\prime}^2 
\Big( \Delta{J}_{x^\prime}^2 - 1 \Big) + 
\Delta{J}_{y^\prime}^2 
\Big( \Delta{J}_{y^\prime}^2 - 1 \Big)\nonumber\\
&& +~~ \frac{1}{2}\Big].
\label{3.5}
\end{eqnarray}
Multiplying and dividing 
$\Delta{J}_{x^\prime}^2$ and $\Delta{J}_{y^\prime}^2$
in the above expression by $2/j$ $(j = \frac{N}{2} = 1)$, we
get
\begin{eqnarray}
S &=& \frac{1}{2}\Bigg[ \frac{2 \Delta{J}_{x^\prime}^2 j}{2j} 
\bigg( \frac{2 \Delta{J}_{x^\prime}^2 j}{2j} - 1 \bigg)\nonumber\\
&& + ~~ 
\frac{2 \Delta{J}_{y^\prime}^2 j}
{2 j} 
\bigg( \frac{2 \Delta{J}_{y^\prime}^2 j}{2 j} - 1 \bigg)
 + \frac{1}{2}\Bigg]\label{3.6}\\
&=& \frac{1}{2}\Bigg[ \frac{ {Q_x}^2 j}{2} 
\bigg( \frac{ {Q_x}^2 j}{2} - 1 
\bigg)\nonumber\\
&& + ~~ 
\frac{ {Q_y}^2 j}
{2} 
\bigg( \frac{ {Q_y}^2 j}{2} - 1 \bigg) + \frac{1}{2}\Bigg],
\label{3.7}
\end{eqnarray}
where $Q_x$ and $Q_y$ are the spin squeezing parameters introduced
in section I. Since $Q_x$ and $Q_y$ are experimentally 
measurable quantities, the parameter $S$ gets connected directly with the experiment. We can obtain numerical values of $S$ by 
measuring $Q_x$ and $Q_y$ by experiment and using the above formula.
  Thus, we can measure $S$ for a system experimentally.

If we now multiply and divide $Q_x^2$ and $Q_y^2$ in Eq. $(\ref{3.7})$ by 
$j/|\langle\hat{\mathbf J}\rangle|^2$, we get
\begin{eqnarray}
S &=& \frac{1}{2}\Bigg[ \frac{ {Q_x}^2 j^2|\langle\hat{\mathbf J}\rangle|^2}{2j
|\langle\hat{\mathbf J}\rangle|^2} 
\bigg( \frac{ {Q_x}^2 j^2|\langle\hat{\mathbf J}\rangle|^2}{2j|\langle\hat{\mathbf J}\rangle|^2} - 1 
\bigg)\nonumber\\
&& + ~~ 
\frac{ {Q_y}^2 j^2|\langle\hat{\mathbf J}\rangle|^2}
{2j|\langle\hat{\mathbf J}\rangle|^2} 
\bigg( \frac{ {Q_y}^2 j^2|\langle\hat{\mathbf J}\rangle|^2}{2j|\langle\hat{\mathbf J}\rangle|^2} - 1
 \bigg) + \frac{1}{2}\Bigg]\nonumber\\
&=& \frac{1}{2}\Bigg[ \frac{\xi_{R_x}^2|\langle\hat{\mathbf J}\rangle|^2}{2j} 
\bigg( \frac{\xi_{R_x}^2|\langle\hat{\mathbf J}\rangle|^2}{2j} - 1 \bigg)\nonumber\\
&& + ~~ 
\frac{\xi_{R_y}^2|\langle\hat{\mathbf J}\rangle|^2}{2j} 
\bigg(\frac{\xi_{R_y}^2|\langle\hat{\mathbf J}\rangle|^2}{2j} - 1 \bigg) + 
\frac{1}{2}\Bigg],
\label{3.8}
\end{eqnarray}
where 
\begin{eqnarray}
\xi_{R_x} &=& \frac{j}{|\langle\hat{\mathbf J}\rangle|} Q_x,
\label{3.9}\\
\xi_{R_y} &=& \frac{j}{|\langle\hat{\mathbf J}\rangle|} Q_y
\label{3.10}
\end{eqnarray}
 are called the spectroscopic squeezing parameters used
in the context of Ramsey spectroscopy \cite{Wineland}. Thus,
the quantum entanglement parameter $S$ gets connected with the
spectroscopic squeezing parameters, which are experimentally 
measurable.

In the next section, we extend these ideas to systems containing $N$ number of two-level atoms.  

\section{Quantum entanglement in a system of $N$ two-level atoms }

An arbitrary symmetric pure state for a system of $N$ two-level atoms in the
$\{ m_1, m_2, m_3,....m_N \}$ representation is given as

\begin{eqnarray}
|\Psi\rangle &=& G_1\bigg\vert\frac{1}{2},\frac{1}{2},....\frac{1}{2}
\bigg\rangle
+ \frac{G_2}{\sqrt{{}^NC_1}}\Bigg[\bigg\vert-\frac{1}{2},\frac{1}{2},\frac{1}{2},....\frac{1}{2}
\bigg\rangle \nonumber\\
&+& \bigg\vert\frac{1}{2},-\frac{1}{2},\frac{1}{2},....\frac{1}{2}
\bigg\rangle +....+ \bigg\vert\frac{1}{2},\frac{1}{2},\frac{1}{2},....
-\frac{1}{2}\bigg\rangle\Bigg]\nonumber\\ 
&+& \frac{G_3}{\sqrt{{}^NC_2}}\Bigg[\bigg\vert-\frac{1}{2},-\frac{1}{2},\frac{1}{2},....\frac{1}{2}
\bigg\rangle \nonumber\\
&+& \bigg\vert-\frac{1}{2},\frac{1}{2},-\frac{1}{2},....\frac{1}{2}
\bigg\rangle +....+ \bigg\vert\frac{1}{2},\frac{1}{2},....
-\frac{1}{2},-\frac{1}{2}\bigg\rangle\Bigg]\nonumber\\
&+& ............+G_{N+1}\bigg\vert-\frac{1}{2},-\frac{1}{2},-
\frac{1}{2},....-\frac{1}{2}\bigg\rangle,
\label{4.1}
\end{eqnarray}

where $G_1$, $G_2$, ..., $G_{N+1}$ are constants and
 ${}^N C_r$ is given as
\begin{equation}
{}^NC_r = \frac{N!}{r!(N-r)!}.
\label{4.1new1}
\end{equation}

The quantum fluctuations $\Delta{J_{x^\prime}^2}$ and
$\Delta{J_{y^\prime}^2}$ for this system can be written in analogy
to Eqs. $(\ref{2.19newm1})$ and $(\ref{2.19newm2})$ as

\begin{eqnarray}
\Delta{J_{x^\prime}^2} &=&
\sum_{i=1}^{N} \Delta{J_{i_{x^\prime}}^2} + \sum_{i=1}^{N}
\sum_{{}^{l=1}_{l\ne i}}^{N}\bigg[\langle\hat{J}_{i_{x}}
\hat{J}_{l_{x}}\rangle\cos^2\theta\cos^2\phi\nonumber\\
&+& \langle\hat{J}_{i_{y}}
\hat{J}_{l_{y}}\rangle\cos^2\theta\sin^2\phi
+ \langle\hat{J}_{i_{z}}
\hat{J}_{l_{z}}\rangle\sin^2\theta\nonumber\\
&+& \langle\hat{J}_{i_{x}}
\hat{J}_{l_y}\rangle\cos^2\theta\sin2\phi
 -\langle\hat{J}_{i_{x}}
\hat{J}_{l_{z}}\rangle\sin2\theta\cos\phi \nonumber\\
&-& \langle\hat{J}_{i_{y}}
\hat{J}_{l_{z}}\rangle\sin2\theta\sin\phi\bigg]
\label{4.2}  
\end{eqnarray}
and
\begin{eqnarray}
\Delta{J_{y^\prime}^2} &=&
\sum_{i=1}^{N} \Delta{J_{i_{y^\prime}}^2} + \sum_{i=1}^{N}
\sum_{{}^{l=1}_{l\ne i}}^{N}\bigg[ \langle\hat{J}_{i_{x}}
\hat{J}_{l_{x}}\rangle\sin^2\phi\nonumber\\
&+& \langle\hat{J}_{i_{y}}
\hat{J}_{l_{y}}\rangle\cos^2\phi
-\langle\hat{J}_{i_{x}}\hat{J}_{l_{y}}\rangle\sin2\phi\bigg],
\label{4.3}
\end{eqnarray}
where the upper index 2 in the summations in Eqs. $(\ref{2.19newm1})$ and $(\ref{2.19newm2})$ has been replaced by $N$.

Now, since the state $|\Psi\rangle$ is symmetric under the exchange of any two atoms and all the atoms have been treated on equal footing,
we have for the state $|\Psi\rangle$,
\begin{eqnarray}
\langle\hat{J}_{1_{x}}\rangle &=& \langle\hat{J}_{2_{x}}\rangle =
......= \langle\hat{J}_{N_{x}}\rangle,\label{4.4}\\
\langle\hat{J}_{1_{y}}\rangle &=& \langle\hat{J}_{2_{y}}\rangle =
......= \langle\hat{J}_{N_{y}}\rangle,\label{4.5}\\
\langle\hat{J}_{1_{z}}\rangle &=& \langle\hat{J}_{2_{z}}\rangle =
......= \langle\hat{J}_{N_{z}}\rangle,\label{4.6}\\
\langle\hat{J}_{1_{x}}\hat{J}_{2_{x}}\rangle &=& 
\langle\hat{J}_{1_{x}}\hat{J}_{3_{x}}\rangle =
......= \langle\hat{J}_{{N-1}_{x}}\hat{J}_{N_{x}}\rangle,\label{4.7}\\
\langle\hat{J}_{1_{y}}\hat{J}_{2_{y}}\rangle &=& 
\langle\hat{J}_{1_{y}}\hat{J}_{3_{y}}\rangle =
......= \langle\hat{J}_{{N-1}_{y}}\hat{J}_{N_{y}}\rangle,\label{4.8}\\
\langle\hat{J}_{1_{z}}\hat{J}_{2_{z}}\rangle &=& 
\langle\hat{J}_{1_{z}}\hat{J}_{3_{z}}\rangle =
......= \langle\hat{J}_{{N-1}_{z}}\hat{J}_{N_{z}}\rangle,\label{4.9}\\
\langle\hat{J}_{1_{x}}\hat{J}_{2_{z}}\rangle &=& 
\langle\hat{J}_{1_{x}}\hat{J}_{3_{z}}\rangle =
......= \langle\hat{J}_{{N-1}_{x}}\hat{J}_{N_{z}}\rangle,
\label{4.10}\\
\langle\hat{J}_{1_{x}}\hat{J}_{2_{y}}\rangle &=& 
\langle\hat{J}_{1_{x}}\hat{J}_{3_{y}}\rangle =
......= \langle\hat{J}_{{N-1}_{x}}\hat{J}_{N_{y}}\rangle,
\label{4.11}\\
\langle\hat{J}_{1_{y}}\hat{J}_{2_{z}}\rangle &=& 
\langle\hat{J}_{1_{y}}\hat{J}_{3_{z}}\rangle =
......= \langle\hat{J}_{{N-1}_{y}}\hat{J}_{N_{z}}\rangle,
\label{4.12}
\end{eqnarray}  
and, also,
\begin{eqnarray}
\Delta{J_{1_{x^\prime}}^2} &=& \Delta{J_{2_{x^\prime}}^2} = ........=
\Delta{J_{N_{x^\prime}}^2} = \frac{1}{4},\label{4.13}\\
\Delta{J_{1_{y^\prime}}^2} &=& \Delta{J_{2_{y^\prime}}^2} = ........=
\Delta{J_{N_{y^\prime}}^2} = \frac{1}{4}.\label{4.14}
\end{eqnarray}
Therefore, using the above equations we can reduce Eqs. $(\ref{4.2})$
and $(\ref{4.3})$ as
\begin{eqnarray}
\Delta{J_{x^\prime}^2} &=&
N \Delta{J_{1_{x^\prime}}^2} + 2\big({}^{N}C_2\big)
\Big[ \langle\hat{J}_{1_{x}}
\hat{J}_{2_{x}}\rangle\cos^2\theta\cos^2\phi\nonumber\\
&+& \langle\hat{J}_{1_{y}}
\hat{J}_{2_{y}}\rangle\cos^2\theta\sin^2\phi + \langle\hat{J}_{1_{z}}
\hat{J}_{2_{z}}\rangle\sin^2\theta\nonumber\\
&+& \langle\hat{J}_{1_{x}}
\hat{J}_{2_{y}}\rangle\cos^2\theta\sin2\phi - \langle\hat{J}_{1_{x}}
\hat{J}_{2_{z}}\rangle\sin2\theta\cos\phi \nonumber\\
&-& \langle\hat{J}_{1_{y}}
\hat{J}_{2_{z}}\rangle\sin2\theta\sin\phi\Big]
\label{4.15}  
\end{eqnarray}
and
\begin{eqnarray}
\Delta{J_{y^\prime}^2} &=&
N \Delta{J_{1_{y^\prime}}^2} + 2\big({}^{N}C_2\big)
\Big[ \langle\hat{J}_{1_{x}}\hat{J}_{2_{x}}\rangle\sin^2\phi
\nonumber\\
&+& \langle\hat{J}_{1_{y}}
\hat{J}_{2_{y}}\rangle\cos^2\phi - \langle\hat{J}_{1_{x}}
\hat{J}_{2_{y}}\rangle\sin2\phi\Big],
\label{4.18}
\end{eqnarray}
respectively.
Now, according to Eqs. $(\ref{1.1a})$, $(\ref{4.4})$, $(\ref{4.5})$, and 
$(\ref{4.6})$,  we have
\begin{eqnarray}
\langle\hat{J}_x\rangle = N \langle\hat{J}_{1_x}\rangle,\label{4.24}\\
\langle\hat{J}_y\rangle = N \langle\hat{J}_{1_y}\rangle,\label{4.25}\\
\langle\hat{J}_z\rangle = N \langle\hat{J}_{1_z}\rangle.\label{4.26}
\end{eqnarray} 
Therefore, using the above three equations and Eqs. $(\ref{2.6})$ and $(\ref{2.7})$, we observe that the expressions of $\cos\theta$ and
$\cos\phi$, in this case, have the same forms as given in 
Eqs. $(\ref{2.19new8})$ and $(\ref{2.19new9})$, respectively. 
Hence, using the expressions of $\cos\theta$, $\cos\phi$, 
$\sin\theta$, and $\sin\phi$ obtained from Eqs. $(\ref{2.19new8})$ and $(\ref{2.19new9})$, and also the expressions of 
$\Delta J_{1_{x^\prime,y^\prime}}^2$
given in Eq. $(\ref{4.13})$ and $(\ref{4.14})$, in 
Eqs. $(\ref{4.15})$ and $(\ref{4.18})$, we get
\begin{eqnarray}
&&\Delta{J}_{x^\prime}^2 = \frac{N}{4} +  
\frac{2({}^NC_2)\langle\hat{J}_{1_z}\rangle^2}{|\langle\hat{\mathbf J}_{1}\rangle|^2\Big(\langle\hat{J}_{1_x}
\rangle^2 + \langle\hat{J}_{1_y}\rangle^2\Big)}\nonumber\\
&\times& \bigg[\langle\hat{J}_{1_x}\hat{J}_{2_x}\rangle
\langle\hat{J}_{1_x}\rangle^2 + 2\langle\hat{J}_{1_x}\hat{J}_{2_y}
\rangle\langle\hat{J}_{1_x}\rangle\langle\hat{J}_{1_y}\rangle
\nonumber\\ 
&+& 
\langle\hat{J}_{1_y}\hat{J}_{2_y}\rangle\langle\hat{J}_{1_y}\rangle^2
\bigg] + \frac{2({}^NC_2)\langle\hat{J}_{1_z}\hat{J}_{2_z}\rangle}
{|\langle\hat{\mathbf J}_{1}\rangle|^2}\bigg[\langle\hat{J}_{1_x}
\rangle^2 + \langle\hat{J}_{1_y}\rangle^2\bigg]\nonumber\\
&-& \frac{4({}^NC_2)\langle\hat{J}_{1_z}\rangle}{|\langle\hat{\mathbf J}_{1}\rangle|^2}\bigg[\langle\hat{J}_{1_x}\hat{J}_{2_z}\rangle
\langle\hat{J}_{1_x}\rangle + \langle\hat{J}_{1_y}\hat{J}_{2_z}\rangle
\langle\hat{J}_{1_y}\rangle \bigg]\label{4.27}\\
&=& \frac{N}{4} + CORRX
\label{4.28}
\end{eqnarray}
and
\begin{eqnarray}
\Delta{J}_{y^\prime}^2 &=& \frac{N}{4} +  
\frac{2({}^NC_2)}{\Big(\langle\hat{J}_{1_x}
\rangle^2 + \langle\hat{J}_{1_y}\rangle^2\Big)}
\bigg[\langle\hat{J}_{1_x}\hat{J}_{2_x}\rangle
\langle\hat{J}_{1_y}\rangle^2 \nonumber\\
&+& \langle\hat{J}_{1_y}\hat{J}_{2_y}
\rangle\langle\hat{J}_{1_x}\rangle^2 - 2\langle\hat{J}_{1_x}
\hat{J}_{2_y}\rangle\langle\hat{J}_{1_x}\rangle\langle\hat{J}_{1_y}
\rangle\bigg]\label{4.29}\\ 
&=& \frac{N}{4} + CORRY
\label{4.30}
\end{eqnarray} 
respectively.

 Here $CORRX$ is the sum of last seven terms on the 
right-hand side of 
Eq. $(\ref{4.27})$ and $CORRY$ is the sum of last three terms on the
right-hand side of Eq. $(\ref{4.29})$, respectively. Thus, we observe from Eqs. $(\ref{4.27})$ and $(\ref{4.29})$ that
the quantum fluctuations $\Delta{J_{x^\prime}^2}$ and
$\Delta{J_{y^\prime}^2}$ of a system of $N$ two-level atoms in an
arbitrary 
symmetric pure state can be obtained by finding out the quatum
fluctuations $\Delta{J_{1_{x^\prime}}^2}$ and
$\Delta{J_{1_{y^\prime}}^2}$ of any single atom and the correlations
among any two atoms only in the assembly.
If the quantum state $|\Psi\rangle$ of the composite system is unentangled, then $|\Psi\rangle$ can be written as a direct product
of the $N$ individual atomic state vectors. In this case, the conditions like those expressed in Eq. $(\ref{2.23})$ are satisfied
and it can be shown that $CORRX$ and $CORRY$ are zero. Therefore, we have,
\begin{eqnarray}
\Delta{J_{x^\prime}^2}\bigg\vert_{un-ent} &=& N \Delta{J_{1_{x^\prime}}^2} = \frac{N}{4}
\label{4.20a1}\\
\Delta{J_{y^\prime}^2}\bigg\vert_{un-ent} &=& N \Delta{J_{1_{y^\prime}}^2} = \frac{N}{4}.
\label{4.20a2}
\end{eqnarray}
Thus, $\Delta{J_{x^\prime}^2}$ and
$\Delta{J_{y^\prime}^2}$ are just the algebraic sum of the quantum fluctuations $\Delta{J_{i_{x^\prime}}^2}$ and
$\Delta{J_{i_{y^\prime}}^2}$ $(i = 1, 2, 3, ....,N)$, respectively,
of all the $N$ individual constituent atoms.
If the quantum state of the composite system is entangled,
the conditions like those given in Eq. $(\ref{2.23})$ are not satisfied and,
hence, $CORRX$ and $CORRY$ are non-zero.
We see that here also $CORRX$ and $CORRY$ are the measures of the deviations of
the quantum fluctuations $\Delta{J_{x^\prime}^2}$ and
$\Delta{J_{y^\prime}^2}$ of an entangled state from
those of an unentangled one. As mentioned in section-II, here also
we construct the quantum entanglement
parameter $S$ as the mean squared deviation of the quantum
 fluctuations in the two quadratures ( $x^\prime$ and $y^\prime$ )
of an entangled state from the corresponding fluctuations of an
unentangled one.
According to Eqs. $(\ref{2.28})$, $(\ref{4.28})$, and $(\ref{4.30})$, we have
\begin{eqnarray}
S &=& \frac{1}{2}\Big[(CORRX)^2 + (CORRY)^2\Big]\nonumber\\
 &=& \frac{1}{2}\Big[
\Delta{J_{x^\prime}}^4 - \frac{N}{2}
\Delta{J_{x^\prime}}^2\nonumber\\
 &+& \Delta{J_{y^\prime}}^4 - \frac{N}{2}
\Delta{J_{y^\prime}}^2 + \frac{N^2}{8}\Big]\nonumber\\
&=& \frac{1}{2}\Big[
\Delta{J_{x^\prime}}^2 \Big( \Delta{J_{x^\prime}}^2  - 
\frac{N}{2}\Big)\nonumber\\
 && + \Delta{J_{y^\prime}}^2\Big(\Delta{J_{y^\prime}}^2  - \frac{N}{2}
\Big) + \frac{N^2}{8}\Big].
\label{4.21}
\end{eqnarray}
  The necessary and
sufficient condition for the presence of quantum entanglement in this
system of $N$ two-level atoms is 
$S > 0$. The proof is as follows. If the composite state vector 
$|\Psi\rangle$ is entangled, it cannot be expressed as a direct product of the $N$ individual atomic state vectors. Then the
 conditions like those given in
Eqs. $(\ref{2.23})$ are not satisfied and, hence, either $CORRX$ or $CORRY$ or both of them
are non-zero, implying that $ S > 0 $. Thus, $S > 0$ forms the 
necessary condition for the presence of entanglement. We now show that
the condition is sufficient also.
 If $S > 0$, either $CORRX$ or $CORRY$ or both of them are non-zero, implying that the conditions like those expressed in Eq. $(\ref{2.23})$ are not all satisfied and, hence, the corresponding quantum state is not expressible as a direct product of the $N$ individual atomic state vectors, implying that the composite state is entangled. Thus,
the condition $S > 0$ forms the sufficient condition for the
presence of entanglement.

As in section-III, we now relate $S$ with experimentally measurable
quantities.
Multiplying and dividing 
$\Delta{J}_{x^\prime}^2$ and $\Delta{J}_{y^\prime}^2$
in Eq. $(\ref{4.21})$ by $2/j$, we
get
\begin{eqnarray}
S &=& \frac{1}{2}\Bigg[ \frac{2 \Delta{J}_{x^\prime}^2 j}{2 j} 
\bigg( \frac{2 \Delta{J}_{x^\prime}^2 j}{2 j} - \frac{N}{2} 
\bigg)\nonumber\\
&& + ~~ 
\frac{2 \Delta{J}_{y^\prime}^2 j}
{2 j} 
\bigg( \frac{2 \Delta{J}_{y^\prime}^2 j}{2 j} - \frac{N}{2} 
\bigg)
 + \frac{N^2}{8}\Bigg]\nonumber\\
\label{3.6}\\
&=& \frac{1}{2}\Bigg[ \frac{ {Q_x}^2 j}{2} 
\bigg( \frac{ {Q_x}^2 j}{2} - \frac{N}{2} 
\bigg)\nonumber\\
&& + ~~ 
\frac{ {Q_y}^2 j}
{2} 
\bigg( \frac{ {Q_y}^2 j}{2} - \frac{N}{2} \bigg) + \frac{N^2}{8}\Bigg],
\label{4.22}
\end{eqnarray}
where $Q_x$ and $Q_y$ are the spin squeezing parameters introduced
in Eqs. $(\ref{1.4})$ and $(\ref{1.5})$, respectively, in section-I. Thus, for a system of $N$ two-level atoms, we can 
measure the quantum entanglement parameter by measuring the spin
squeezing parameters $Q_x$ and $Q_y$ of the system.

If we now multiply and divide $Q_x^2$ and $Q_y^2$ in Eq. $(\ref{4.22})$ by $j / |\langle\hat{\mathbf J}\rangle|^2$,  we get
\begin{eqnarray}
S &=& \frac{1}{2}\Bigg[ \frac{ {Q_x}^2 j^2|\langle\hat{\mathbf J}\rangle|^2}{2j
|\langle\hat{\mathbf J}\rangle|^2} 
\bigg( \frac{ {Q_x}^2 j^2|\langle\hat{\mathbf J}\rangle|^2}{2j|\langle\hat{\mathbf J}\rangle|^2} - \frac{N}{2} 
\bigg)\nonumber\\
&& + ~~ 
\frac{ {Q_y}^2 j^2|\langle\hat{\mathbf J}\rangle|^2}
{2j|\langle\hat{\mathbf J}\rangle|^2} 
\bigg( \frac{ {Q_y}^2 j^2|\langle\hat{\mathbf J}\rangle|^2}{2j|\langle\hat{\mathbf J}\rangle|^2} - \frac{N}{2}
 \bigg) + \frac{N^2}{8}\Bigg]\nonumber\\
&=& \frac{1}{2}\Bigg[ \frac{\xi_{R_x}^2|\langle\hat{\mathbf J}\rangle|^2}{2j} 
\bigg( \frac{\xi_{R_x}^2|\langle\hat{\mathbf J}\rangle|^2}{2j} - 
\frac{N}{2} \bigg)\nonumber\\
&& + ~~ 
\frac{\xi_{R_y}^2|\langle\hat{\mathbf J}\rangle|^2}{2j} 
\bigg(\frac{\xi_{R_y}^2|\langle\hat{\mathbf J}\rangle|^2}{2j} - 
\frac{N}{2} \bigg) + 
\frac{N^2}{8}\Bigg],
\label{4.23}
\end{eqnarray}

where $\xi_{R_x}$ and $\xi_{R_y}$ are the spectroscopic squeezing
parameters \cite{Wineland} already introduced in section III.
 Thus, we relate the quantum entanglement parameter
$S$ of a system of $N$ two-level atoms with the experimentally 
measurable squeezing parameters used in the Ramsey spectroscopy.

\section{Summary and Conclusion}

We proposed the necessary and sufficient condition for the presence of quantum entanglement in arbitrary symmetric pure states of two-level atoms. We took the quantum fluctuations of the system in terms
of the components of the pseudo-spin vector operator 
$\hat{\mathbf J}$ in two mutually orthogonal
directions in a plane perpendicular to the mean pseudo-spin vector
$\langle\hat{\mathbf J}\rangle$. We then expressed these fluctuations
as an algebraic sum of the fluctuations of the individual constituent
atoms and their correlation terms. We took these correlation terms in the two mutually orthogonal directions and in a plane
perpendicular to $\langle\hat{\mathbf J}\rangle$ to construct a 
parameter $S$, called the quantum entanglement parameter. We showed
that this parameter can be used to detect and quantify quantum
entanglement. The necessary and sufficient condition for the presence
of quantum entanglement in such systems is $S > 0$. If a quantum state of the system is unentangled, we have $S = 0$. We also said that since $S$ is the mean squared deviation of the quantum fluctuations of an entangled state from the corresponding fluctuations of an unentangled one, the numerical value of $S$ can be taken as a measure of quantum entanglement in the system. We first made all these studies in case of two two-level atoms
and then extended these ideas in case of systems containing $N$ number
of such atoms. We also established the relationship between the
quantum entanglement parameter $S$ and spin squeezing and 
spectroscopic squeezing parameters. This shows how we can measure $S$
experimentally. We hope that our study may produce deeper insight into
the subject.

\section{Acknowledgements}

I am grateful to Nilakantha Nayak and Biplab Ghosh for useful 
discussions. I am also grateful to Sanjit Kumar Das for helping me solve the technical problems in my personal computer while preparing this manuscript. I am thankful to Sunandan Gangopadhyay for helping me to use some features of LATEX program for the preparation of this manuscript. I am also grateful to 
Archan S. Majumdar for endorsing me to become an arXiv user in
quant-ph section.

\end{document}